\documentclass[12pt]{article}
\usepackage{epsf}
\usepackage{alltt}
\def\largelinestretch{\renewcommand{\baselinestretch}{1.1}}
\largelinestretch\small\normalsize

\title{
\vspace*{15mm}
{\bf On the Charge Asymmetry of the Like-Sign Lepton Pairs
     Induced by $B-\overline{B}$ Production Asymmetry}
}

\author{
        {\it A.~Belkov, T.~Ilitcheva, 
             S.~Shulga\footnote{Permanent address: Francisk Skarina Gomel 
                                State University, Belarus}}\\
        JINR, Dubna
}
\date{ }
 
\begin{document}
\maketitle
\begin{abstract}
  In Monte Carlo simulation of $pp$ and $pn$ interactions, it is shown that 
the charge asymmetry of like-sign lepton pairs can be observed as a 
manifestation of $B-\overline{B}$ production asymmetry.
  In this way, the $B-\overline{B}$ production asymmetry could be studied 
experimentally without full reconstruction of $B$-mesons.
\end{abstract}

   It is known that the asymmetry between decays of $B$ and $\bar{B}$ 
mesons provides a signal for CP violation, but can also be caused by the 
production asymmetry between $B$ and $\overline{B}$ mesons in $pp$ or $pn$
collisions.
   The physics origin of asymmetry of heavy-meson production in hadronic
interactions is related to a simple effect of the valence quarks in the 
colliding hadrons and non trivial dynamics of the hadronization process 
\cite{sjostrand,ingelman}.

\begin{figure}[hbt]
\begin{center}
\epsfxsize=.6\textwidth\epsfbox{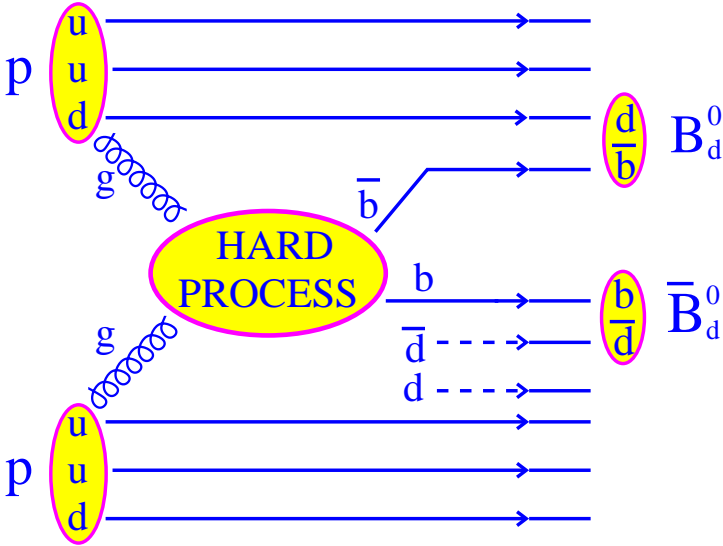}
\caption{\small The mechanism of the production asymmetry in $pp$ collisions
                due to asymmetry in flavours of quarks and antiquarks with 
                which the produced $b$ and $\bar{b}$ quarks can combine to 
                form $B$ mesons
}
\vspace*{-5mm}
\label{frg_asymm}
\end{center}
\end{figure}

   At the parton level, the $b$ and $\bar{b}$ quarks are generated 
symmetrically since their production is described within perturbative QCD by 
diagrams of hard scattering of partons which always arise through the 
$g\to b\bar{b}$ or $gg\to b\bar{b}$ couplings.
   As the result of the strong interaction in the confinement region of QCD, 
the colored quarks produced perturbatively in the hard scattering processes 
are transformed into colorless hadrons.
   In the case of $b$ quarks, this transformation (quark hadronization) may
introduce an asymmetry between a $B$ meson and its antiparticle if there is an
asymmetry in the quark and antiquark flavours that are available in the 
remnants of the initial hadron for $B$-meson formation (see 
Fig.~\ref{frg_asymm}).

   Quark hadronization includes non-perturbative QCD processes and is 
treated and simulated by PYTHIA package \cite{PYTHIA} in the frame of the 
string fragmentation model.
   The Monte Carlo studies by using PYTHIA (see Ref.~\cite{ingelman}) show 
that the overall $B-\overline{B}$ production asymmetry is expected to be below
one percent for $pp$ collisions at LHC and becomes larger at smaller energies,
giving a few percent or even above ten percent for $pp$ and $pn$ interactions 
at HERA-B.
   Thus, an experimental study of the $B-\overline{B}$ production asymmetry
should therefore be most feasible at HERA-B with proton beam of 920 GeV.
   The measurement of the asymmetry in the restricted phase space regions, i.e.
as a function of transverse momentum, rapidity, pseudorapidity, or Feynman 
variable $x_F$ of $B$ mesons, provides much more information on the 
non-perturbative dynamics of quark hadronization.
   According to Monte Carlo studies of Ref.~\cite{ingelman}, the 
$B-\overline{B}$ production asymmetry is positive and of order 10\% in the 
central region of phase space, but becomes negative and very large in the 
extreme forward/backward regions close to the remnants.

   In spite of the great importance of a better understanding of 
$B-\overline{B}$ asymmetry for CP-violation studies and as a test of 
hadronization models as well as a mean to constrain their parameters, no 
measurements of this effect in hadronic $B$-meson production has yet been 
performed.
   The direct measurement of the $B-\overline{B}$ asymmetry requires a 
triggering and reconstruction of $B$-decays.
   One of the possible ways to perform the asymmetry measurement is to use 
lepton pairs from doubly semileptonic decays $B\to l^+ X$ and 
$\bar{B}\to l^- X$ to trigger the $B$-meson signal.
   The main difficulties in this case are related with the high rate of  
background from leptons coming from decays of charmed particles, resonances, 
kaons and pions.
   A separation of $e\mu$ pairs avoids resonance decays, Drell Yan production 
and other processes which feed di-electron and di-muon channels. 
   But even after applying the additional kinematical cuts the problem of 
background reduction is not completely solved because of the dominance of 
the $e\mu$ signal from the doubly semileptonic decays of charmed particles, 
mainly from decays $D^{(*)}\to l^+ X$ and $\bar{D}^{\,(*)}\to l^- X$. 

   In this paper we propose to trigger the $B$-meson signal by selection of 
the like-sign lepton pairs $l^\pm l^\pm$.
   This makes it possible to largely suppress the background originating from 
decays of charm particles and, after applying an additional cut on lepton 
transverse momentum, the $l^\pm l^\pm$ signal from doubly semileptonic decays 
of $B$ mesons becomes clearly seen.
   Moreover, the $B-\overline{B}$ production asymmetry induces the charge
asymmetry in like-sign lepton pairs which, therefore, reflects the 
non-perturbative dynamics of hadronization processes but can be measured 
directly without reconstruction of $B$-meson decays.
   The detailed Monte Carlo studies of lepton pair production in $pp$ 
interactions at the proton energy of HERA-B have been performed for this paper
by using the version PYTHIA 6.158.
   We analyze the contribution of various physical sources to like-sign lepton
pair production and their asymmetries.
   A special study has been devoted to the role of $B^0-\overline{B}^0$ 
oscillations which obscure the lepton-pair asymmetry induced by asymmetry of
$B-\overline{B}$ production (see also discussions in Ref.~\cite{ingelman}).
  
\begin{figure}[hbt]
\begin{center}
\epsfxsize=.9\textwidth\epsfbox{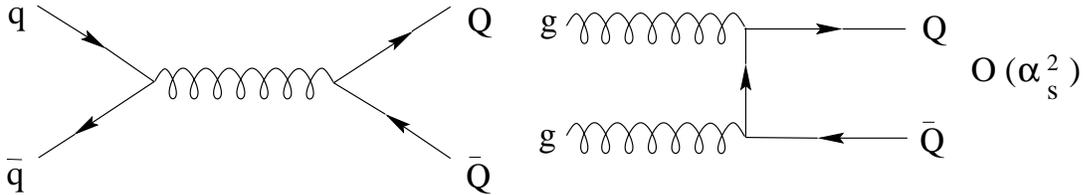}
\end{center}
\caption{\small Feynman diagrams for direct production of heavy quarks 
                ($Q$ is $c$ or $b$ quarks) via parton fusion mechanisms}
\label{pfQ}
\end{figure}

   In this paper we consider only $pp$ interactions at fixed target with 
proton beam energy of 920 GeV.
   Generation of $b\bar{b}$ and $c\bar{c}$ events in PYTHIA is based on the 
description of heavy flavor production within the usual parton model.
   At HERA-B energy, this approach assumes that light partons in the incoming 
protons collide and produce pairs of heavy quarks $Q \bar{Q}$ 
($Q$ denotes $c$ or $b$ quark) predominantly via the hard scattering processes
of parton fusion $q\bar{q} \to Q\bar{Q}$ or $gg \to Q\bar{Q}$ corresponding
to the lowest-order (leading) graphs shown in Fig.~\ref{pfQ}.
   The next-to-leading order graphs of flavor excitation and gluon splitting
mechanisms gain in importance only as the c.~m.~energy is considerably
increased.

   Therefore, we use PYTHIA 6.158 with the option {\tt MSEL}~=~5 
to generate the $b\bar{b}$ events and {\tt MSEL}~=~4 for production of
$c\bar{c}$ pairs ({\tt MSEL} is a steering parameter in the {\tt PYSUBS}
common block).
   In this regime the simulation of heavy flavor production is performed only
via the parton fusion mechanism with massive matrix elements for quark 
generation. 
   Each event contains at least one $Q\overline{Q}$ pair.
   The Gl\"uck-Reya-Vogt (GRV94L) leading order proton parton distribution set 
and SLAC (Peterson) fragmentation function have been used, which are available
when setting {\tt MSTP(51)}=4 (by default) and {\tt MSTJ(11)}=3, respectively. 
   By default, the mechanism of neutral $B$-meson oscillations is switched on,
which corresponds to setting MSTJ(26)=2 with oscillation parameters 
{\tt PARJ(76)}~=~$x_d$~=~0.7 and {\tt PARJ(77)}~=~$x_s$~=~20.

\begin{figure}[hbt]
\begin{center}
\begin{tabular}{c}
\epsfxsize=.85\textwidth\epsfbox{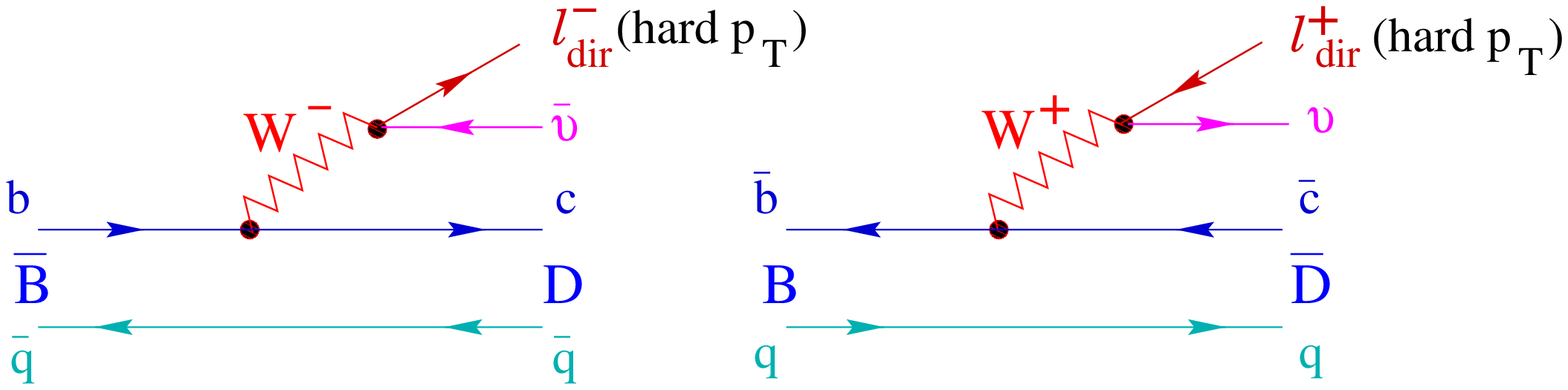} \\
a) \\ \\
\epsfxsize=.7\textwidth\epsfbox{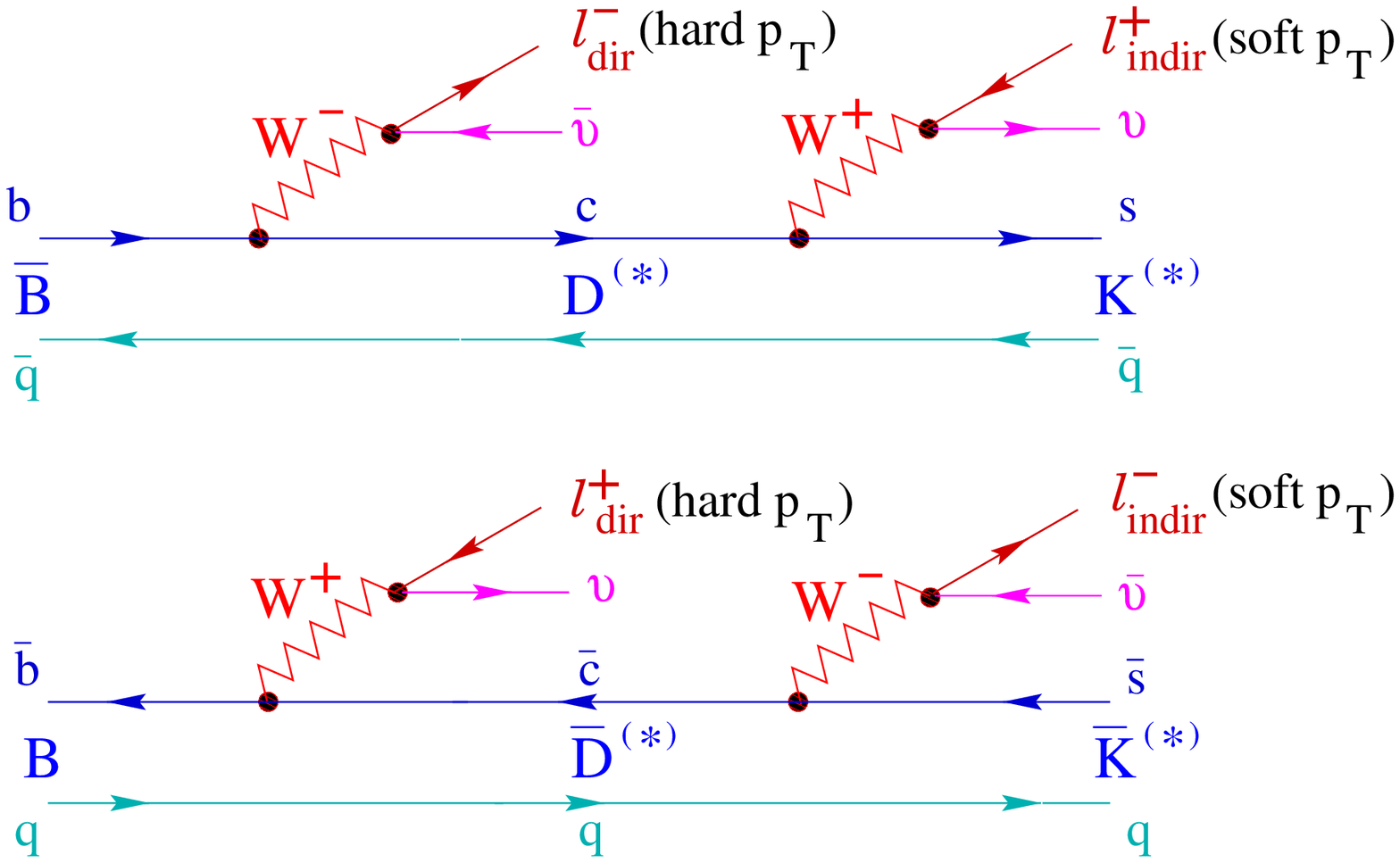} \\ 
b) \\ \\
\epsfxsize=.85\textwidth\epsfbox{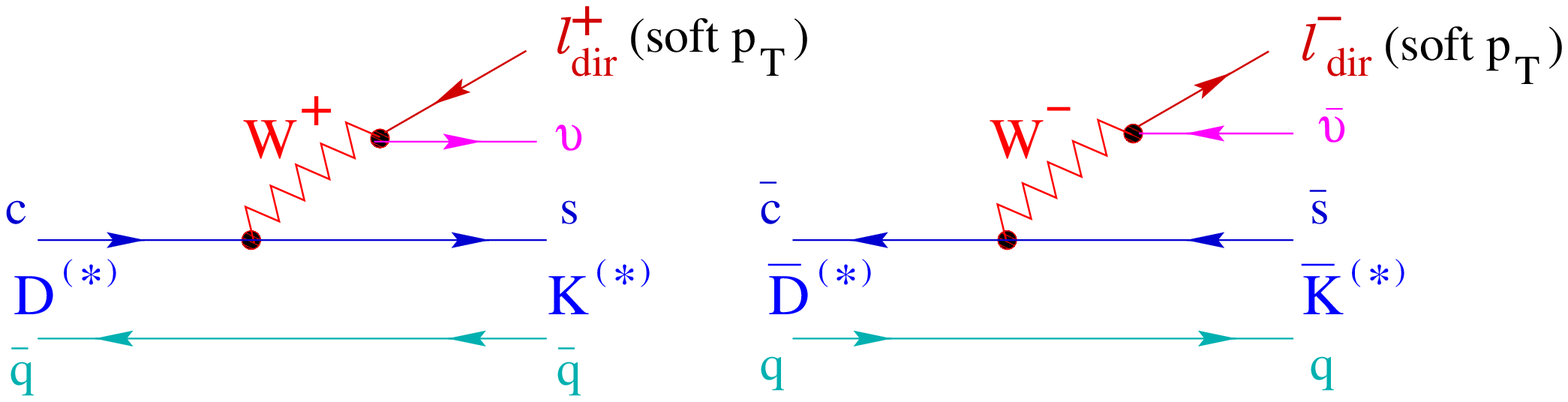}\\
c) \\
\end{tabular}
\end{center}
\caption{\small Classification of leptons induced by semileptonic decays of 
                heavy quarks: 
                a) direct leptons from $b$ or $\bar{b}$ due to transitions 
                   $b\to l^- X$ or $\bar{b}\to l^{\prime +} X^\prime$;
                b) indirect (cascade) leptons from $b$ or $\bar{b}$ due to
                   transitions $b\to l^- Xc(\to l^{\prime +} X^\prime)$ or
                   $\bar{b}\to l^+ X\bar{c}(l^{\prime -} X^\prime )$
                c) direct leptons from $c$ or $\bar{c}$ due to transitions
                   $c\to  l^+ X$ or $\bar{c}\to l^{\prime -} X^\prime$
}
\label{lepton_class}
\end{figure}

   Let us distinguish three types of leptons originating from semileptonic weak
transitions of heavy quarks -- direct and indirect leptons from $b\bar{b}$ 
events, and direct leptons from $c\bar{c}$ events -- according to the 
classification illustrated in Fig.~\ref{lepton_class}.
   In Fig.~\ref{lepton_class} only specific decays of heavy mesons are shown 
but a full sample of events generated by PYTHIA contains also leptons
originating from decays of heavy quark in baryons as well as decays of other 
particles.

   The rates of various types of $b\bar{b}$ and $c\bar{c}$ events with charged
leptons in final state, estimated using PYTHIA, are shown in 
Table~\ref{lept_rates}.
   The ratios of cross sections of $c\bar{c}$ and $b\bar{b}$ leptonic events 
were calculated assuming a ratio of total cross sections 
$\sigma^{tot}_{c\bar{c}}/\sigma^{tot}_{b\bar{b}}\approx 1000$.
   The leptons in the final state of $b\bar{b}$ and $c\bar{c}$ events can 
arise from both decays of heavy mesons or baryons and other sources including
decays of kaons and pions.
   Table~\ref{lept_rates} shows that the $c\bar{c}$ background dominates not 
only in the case of $e\mu$-pair selection but even for events with like-sign 
lepton pairs $l^\pm l^\pm$ if no cut on lepton transverse momentum $p_T$ has 
been applied.
   A significant relative decrease of $c\bar{c}$ background in the case of 
$l^\pm l^\pm$ pair selection as compared with $e\mu$ pairs is caused by a fact
that the contribution of doubly semileptonic decays of $b\bar{b}$ to 
$l^\pm l^\pm$ sample arises as a combination of direct and indirect leptons,
\begin{equation}
b\to l^-X~~\&~~\bar{b}\to l^+X\bar{c}(\to l^{\prime -}X^\prime)
~~~\mbox{or}~~~
b\to l^-Xc(\to l^+X)~~\&~~\bar{b}\to l^{\prime -}X^\prime\,,
\label{110}
\end{equation}
and combination of direct leptons due to $B^0-\bar{B}^0$ oscillations,
\begin{equation}
b\to l^-X~~\&~~\bar{b}\to b\to l^-X
~~~\mbox{or}~~~
b\to \bar{b}\to l^{\prime -}X^\prime~~\&~~\bar{b}\to l^{\prime -}X^\prime\,,
\label{200}
\end{equation}
while the contribution of $c\bar{c}$ events to the $l^\pm l^\pm$ sample arises 
mainly as a combinatorial background involving decays of kaons and pions.

\begin{table}
\caption{Fractions of events with charged leptons without and with applying a
         cut on lepton transverse momentum $p_T$ for various types of 
         event selection: a) events with one or more charged leptons; 
         b) $\mu^{\pm}e^{\mp}$ pairs; c) like-sign lepton pairs $l^\pm l^\pm$}
\label{lept_rates}
\begin{center}
\begin{tabular}{|l|l|l|l|}\hline\hline
Event     & Values & No~$p_T$~cut & $p_T\geq 1$ GeV\\
selection &        &                 &
\\ \hline\hline
   & $\sigma^{(l\geq1)}_{b\overline{b}}$/$\sigma^{tot}_{b\overline{b}}$
   & 0.63 & 0.26 \\[2mm]
a) & $\sigma^{(l\geq1)}_{c\overline{c}}$/$\sigma^{tot}_{c\overline{c}}$
   & 0.34 & 0.012 \\[2mm]
   & $\sigma^{(l\geq1)}_{c\overline{c}}$/$\sigma^{(l\geq1)}_{b\overline{b}}$
   & 540  & 45\\[2mm]\hline

   &
   $\sigma^{\mu^{\pm}e^{\mp}}_{b\overline{b}}$/$\sigma^{tot}_{b\overline{b}}$
   & 0.089 & $8.9\times 10^{-3}$ \\[2mm]
b) &
$\sigma^{\mu^{\pm}e^{\mp}}_{c\overline{c}}$/$\sigma^{tot}_{c\overline{c}}$ 
   & 0.019 & $2.8\times 10^{-5}$ \\[2mm]
&   $\sigma^{\mu^{\pm}e^{\mp}}_{c\overline{c}}$/$\sigma^{\mu{\pm}e{\mp}}_{b\overline{b}}$
   & 210  & 3.2\\[2mm]\hline

   &
   $\sigma^{(l^{\pm}l^{\pm})}_{b\overline{b}}$/$\sigma^{tot}_{b\overline{b}}$
   & 0.096 & $6 \times 10^{-3}$ \\[2mm]
c) &
$\sigma^{(l^{\pm}l^{\pm})}_{c\overline{c}}$/$\sigma^{tot}_{c\overline{c}}$ 
   & 0.0014 & $6 \times 10^{-8}$ \\ [2mm]
&
   $\sigma^{(l^{\pm}l^{\pm})}_{c\overline{c}}$/$\sigma^{(l^{\pm}l^{\pm})}_{b\overline{b}}$
   & 15  & 0.01\\[2mm]\hline
\end{tabular}
\end{center}
\end{table}

   Due to simple kinematical arguments the lepton transverse momentum $p_T$ is
larger for leptonic decays of heavier particles.
   Therefore, one can expect that the $p_T$ cut can significantly reduce the 
combinatorial $c\bar{c}$ background to the signal of doubly semileptonic 
decays of $b\bar{b}$, especially, in the case of selection of like-sign lepton
pairs.
   Table~\ref{lept_rates} shows such a reduction of the $c\bar{c}$ background 
and its complete elimination in the case of $l^\pm l^\pm$-pairs selection 
after applying the requirement $p_T\geq 1$ GeV.
   Thus, the selection of like-sign lepton pairs with the additional $p_T$ cut
can be efficiently used to trigger the signal of doubly semileptonic decays 
induced by decays of $b\bar{b}$.
   Therefore, to study the asymmetry of lepton pairs induced by 
$B-\overline{B}$ production asymmetry, we restrict ourself to consideration 
only of the $l^\pm l^\pm$ pairs.

   Let us define the overall charge asymmetry of like-sign lepton pairs as
\begin{equation}
   A = \frac{N(l^+l^+)-N(l^-l^-)}{N(l^+l^+)+N(l^-l^-)}\,.
\label{asym}
\end{equation}
   For Monte Carlo studies, a sample of more then $2\times 10^6$ $b\bar{b}$ 
events with $l^\pm l^\pm$ pairs in final state with $p_T \ge 1$ GeV has been 
generated by PYTHIA.
   The prehistory of each particle from an event can be derived from the
information stored in {\tt PYJETS} common block, and the origin of each lepton
can be traced back to the parton level, i.e. to the decay of heavy quark 
inducing the lepton.
   In this way, in accordance with the classification of leptons given in 
Fig.~\ref{lepton_class}, each event can be characterized by a code number 
``IJK'', where ``I'' is the number of direct leptons from $b\bar{b}$, ``J'' is
the number of indirect leptons from $b\bar{b}$, and ``K'' is the number of 
leptons from other quarks.

   There are two types of events dominating in the generated sample with
$b\bar{b}$-induced $l^\pm l^\pm$ pairs corresponding with the following code 
numbers:
\begin{itemize}
\item ``110'' -- with one direct and one indirect lepton originated from
               doubly semileptonic decays (\ref{110}) of $b\bar{b}$;
\item ``200'' -- contribution of $B^0-\overline{B}^0$ oscillations (\ref{200}).
\end{itemize}
   The full Monte Carlo sample also contains a fraction of events (about 
20\% in total) with the code numbers ``210'', ``120'', ``020'', ``101'' and
``011''.
   The event statistics and results on the estimates of overall asymmetry 
(\ref{asym}) for the full sample of $b\bar{b}$-induced like-sign lepton pairs 
and the dominating fraction of ``110'' and ``200'' events are shown in
Table~\ref{stat_frac_asym} for various $p_T$ cuts.
   The fraction of the ``110'' events is decreased while the fraction of the 
$B^0-\overline{B}^0$ oscillation contribution ``200'' is increased when 
increasing the $p_T$ cut.

\begin{table}
\caption{Total Monte Carlo statistics, fraction of ``110'' and ``200'' events,
         and overall charge asymmetries (\ref{asym}) for $b\bar{b}$-induced 
         like-sign lepton pairs for various $p_T$ cuts}
\label{stat_frac_asym}
\begin{center}
\begin{tabular}{|c|c|c|c|c|c|c|}
\hline
 $p_T$ cut &   Number  & \multicolumn{2}{|c|}{Fraction (\%)} 
                       & \multicolumn{3}{|c|}{Overall asymmetry (\%)}
\\ \cline{3-7}
    (GeV)  & of events & ``110'' & ``200'' & Total & ``110'' & ``200'' \\ 
\hline
1 & 3~251~390 & 43 & 41 & $2.00\pm 0.06$ & $-0.99\pm 0.09$ & $5.79\pm 0.11$ \\
2 &   128~987 & 27 & 70 & $3.55\pm 0.27$ & $-1.92\pm 0.63$ & $5.84\pm 0.44$ \\
3 &     5~128 & 19 & 78 & $2.22\pm 1.4 $ & $ 2.8 \pm 3.6 $ & $0.5 \pm 1.7 $ \\ 
\hline
\end{tabular}
\end{center}
\end{table}

   The main source of $b\bar{b}$ events contain decays of $B$-mesons while the
contribution of $b$-baryon decays does not exceed $1\%$.
   Therefore the contribution of the $b$-baryon production asymmetry to the
total charge asymmetry of lepton pairs is small, and the latter one reflects 
mainly the $B-\overline{B}$-production asymmetry.
   If we define the overall $B-\overline{B}$-production asymmetry as
\begin{equation}
   A(B_q) = \frac{N(B^{(\bar{b}q)})-N(\overline{B}^{(b\bar{q})})}
                 {N(B^{(\bar{b}q)})+N(\overline{B}^{(b\bar{q})})}\,,
\end{equation}
the simulation of $B$-meson production in $pp$-interactions at the HERA-B
energy gives the following estimates:
$$
A(B^0_d) \approx  -0.3\%\,,\qquad
A(B^\pm) \approx   4.2\%\,,\qquad
A(B^0_s) \approx -10.2\%\,.\qquad 
$$

\begin{figure}[hbt]
\begin{center} 
\begin{tabular}{cc}
\epsfxsize=.5\textwidth\epsfbox{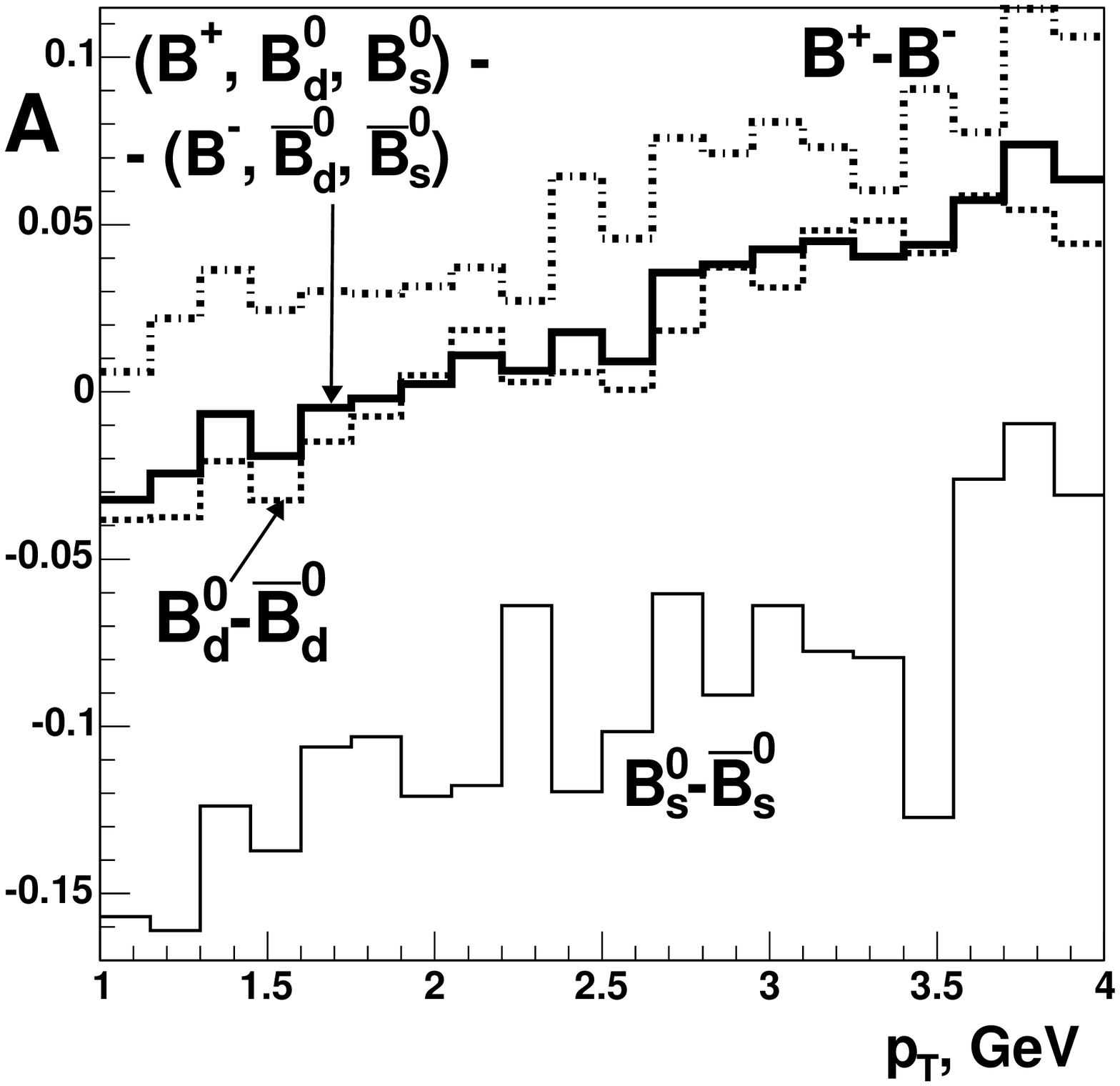} &
\epsfxsize=.5\textwidth\epsfbox{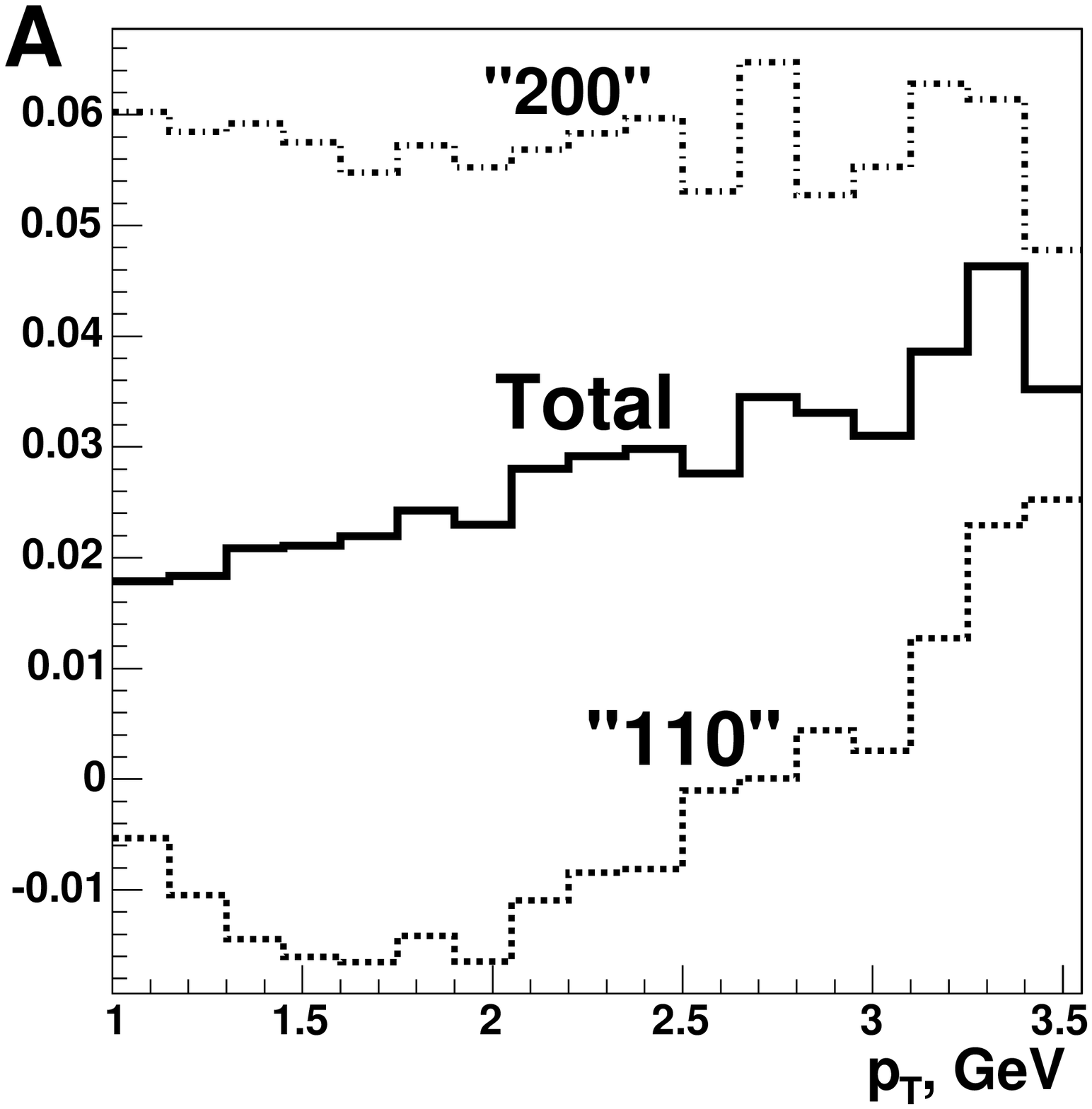}
\end{tabular}
\end{center}
\vspace*{-3mm}
\caption{\small Dependence of charge asymmetry of (a) B-mesons
               and (b)  $b\bar{b}$-induced
                $l^\pm l^\pm$ pairs  on the lepton transverse momentum $p_T$
}
\label{pT_distr}
\end{figure}

\begin{figure}[hbt]
\begin{center} 
\begin{tabular}{cc}
\epsfxsize=.5\textwidth\epsfbox{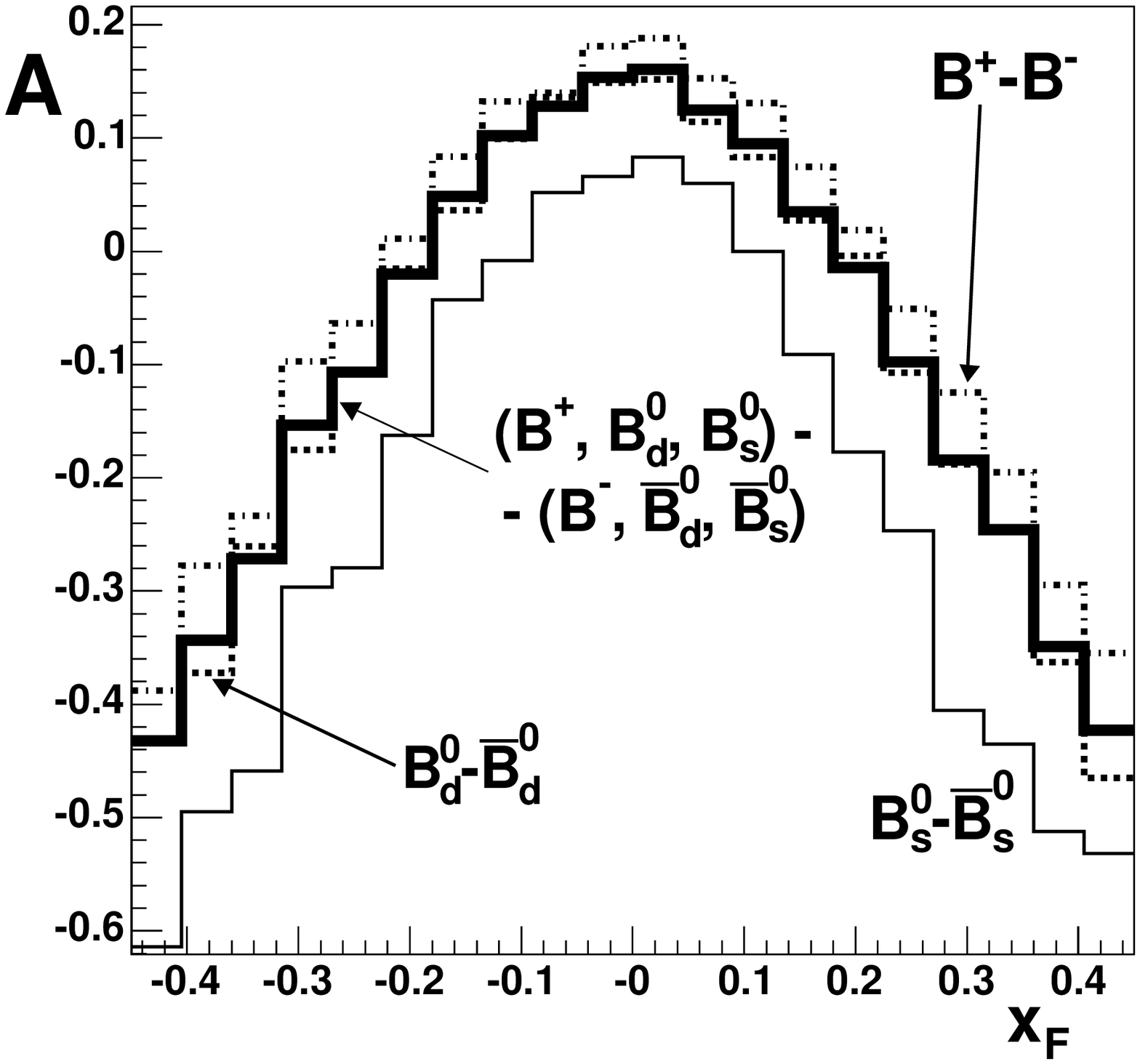} &
\epsfxsize=.5\textwidth\epsfbox{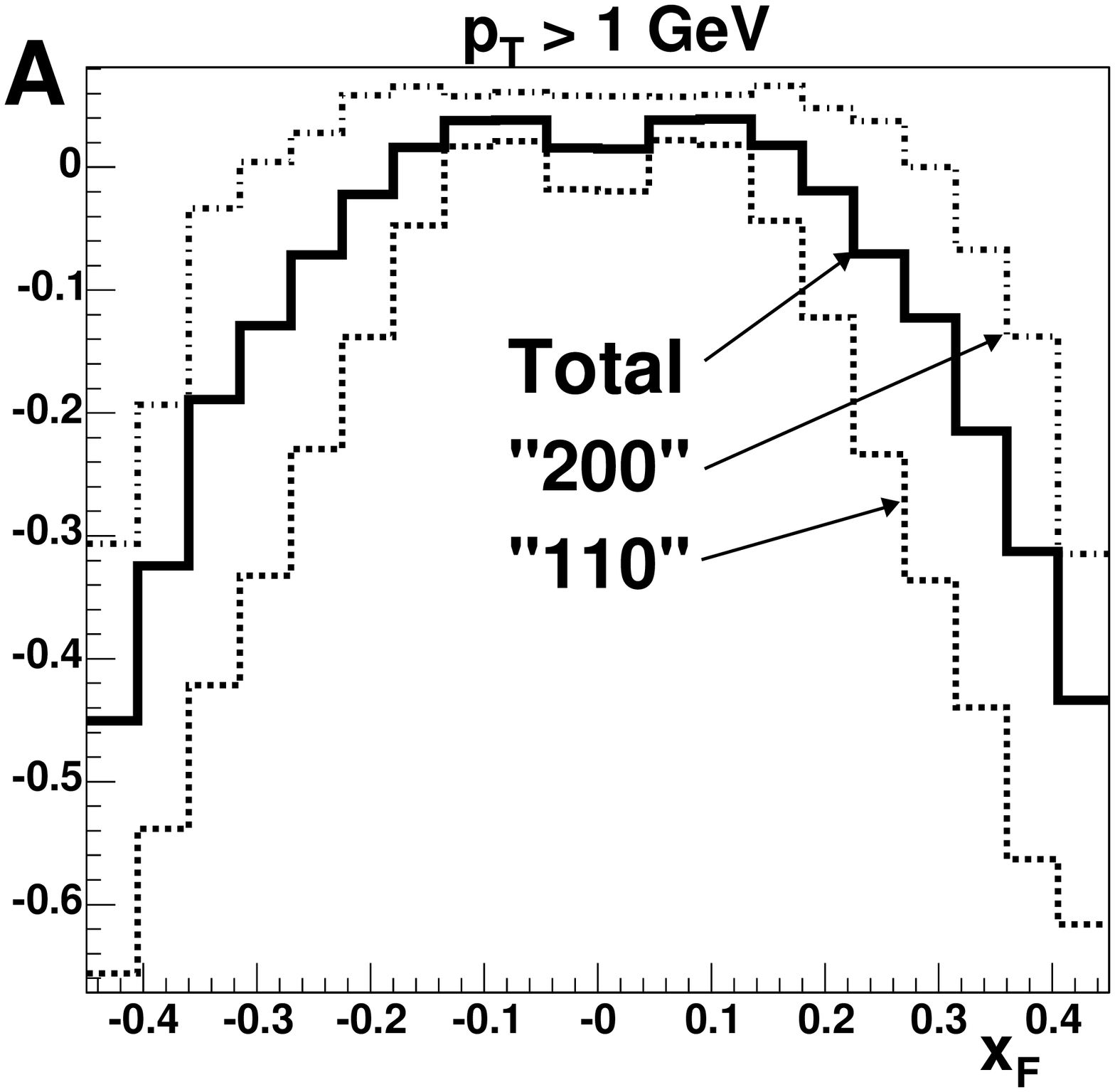}
\end{tabular}
\end{center}
\vspace*{-3mm}
\caption{\small Dependence of charge asymmetry of (a) B-mesons and
                (b) $b\bar{b}$-induced
                $l^\pm l^\pm$ events  on the Feynman variable $x_F$ }
\label{xF_distr}
\end{figure}

\begin{figure}[hbt]
\begin{center} 
\begin{tabular}{cc}
\epsfxsize=.5\textwidth\epsfbox{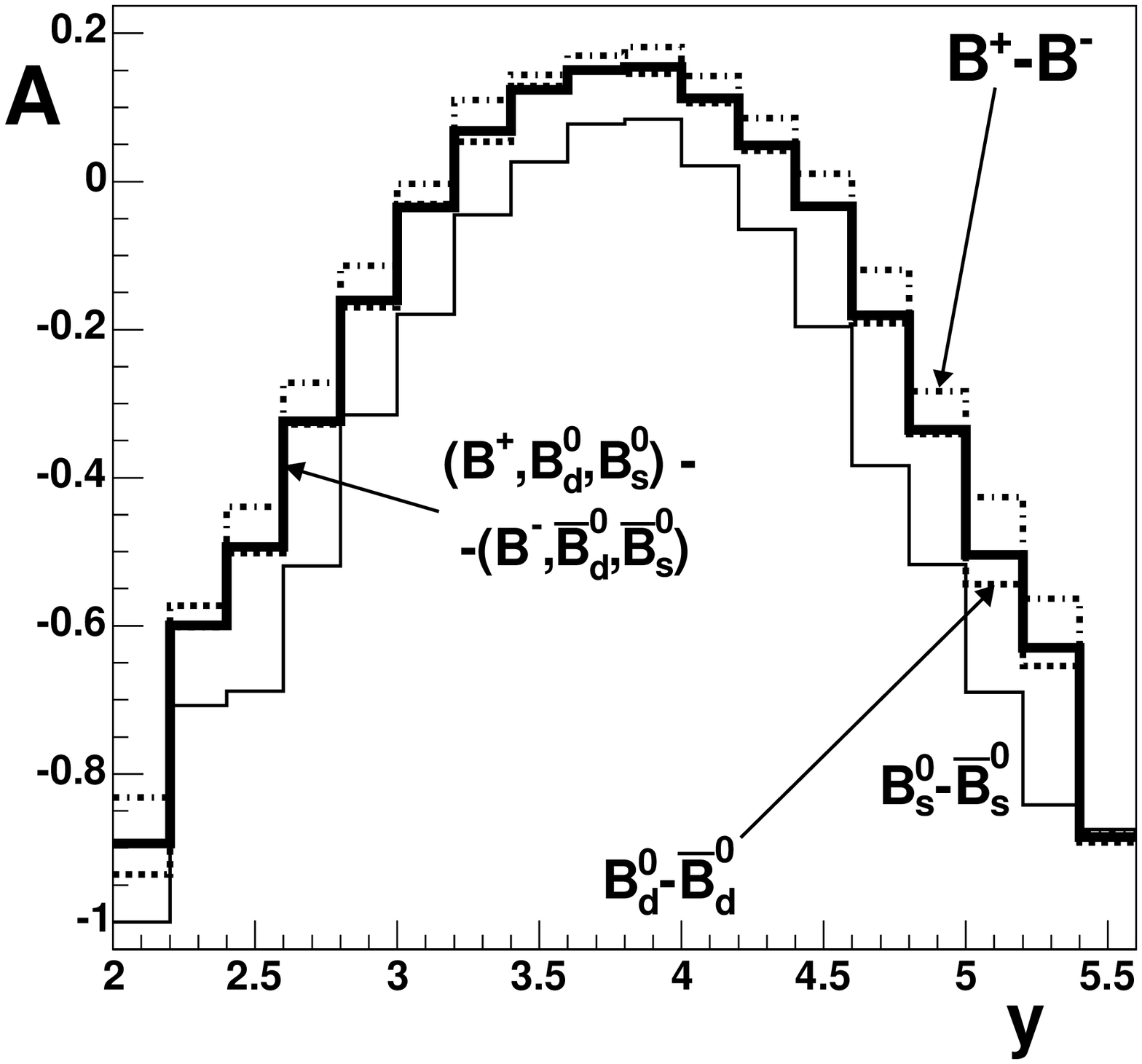} &
\epsfxsize=.5\textwidth\epsfbox{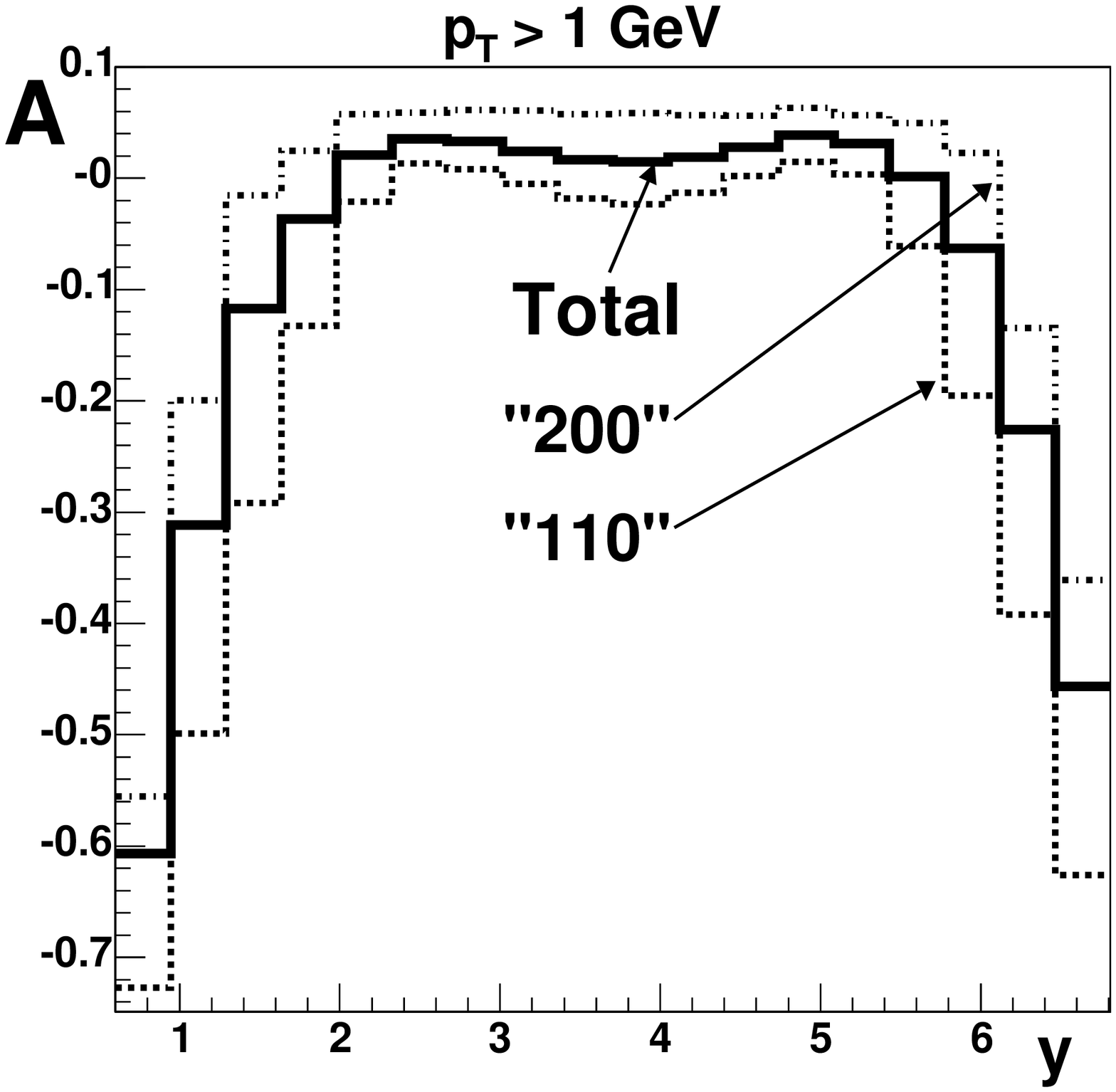}
\end{tabular}
\end{center}
\vspace*{-3mm}
\caption{\small Dependence of charge asymmetry of (a) B-mesons and
                (b) $b\bar{b}$-induced
                $l^\pm l^\pm$ events  on the rapidity
                $y$ }
\label{y_distr}
\end{figure}

   The lepton charge asymmetries in the restricted phase space regions, i.e. 
as a functions of lepton transverse momentum $p_T$, Feynman variable $x_F$
and rapidity $y$ of $l^\pm l^\pm$ pairs, are shown in Figs.~\ref{pT_distr},
\ref{xF_distr}, and \ref{y_distr}.
   Within errors due to limited Monte Carlo statistics, there is no noticeable
$p_T$ dependence in the lepton charge asymmetry of the $l^\pm l^\pm$ events 
with the code number ``200'', caused by $B^0-\overline{B}^0$ oscillations.
   The considerable $p_T$ dependence in the lepton charge asymmetry of total 
sample of events reflects the strong dependencies of the contribution of 
``110'' events on this kinematical parameter.
   The shapes of the corresponding histograms for ``110'' events in 
Figs.~\ref{pT_distr}, \ref{xF_distr}, and \ref{y_distr} reproduce 
qualitatively the $p_T$, $x_F$ and $y$ dependences of $B-\overline{B}$ 
production asymmetries.

   To ensure that the observed effect of the charge asymmetry of the
like-sign lepton pairs is really a manifestation of the $B-\overline{B}$ 
production asymmetry, we have repeated for the case of heavy-quark production 
in $p\bar{p}$ collisions a similar analysis of like-sign lepton pairs with 
the same selection criteria. 
   Because of beam remnant symmetry in $p\bar{p}$ interactions, there is no 
$B-\overline{B}$ production asymmetry, and no charge asymmetry of lepton pairs
was observed.

   Summarizing, we have found that at the proton energy $E =920$ GeV, HERA-B 
provides a unique opportunity to study  the $B-\overline{B}$ production 
asymmetry, caused at the fragmentation level by effects of asymmetric beam 
remnants for $b$ and $\bar{b}$ quarks, by direct measurements of charge 
asymmetry in the production of like-sign lepton pairs without reconstruction 
of $B$-meson decays.
   Our estimates shows that during a one year data taking run the statistics 
on $l^\pm l^\pm$ events at HERA-B would be large enough to provide a
statistical error on charge asymmetry measurements at a level of a few 
percent even in the bins at the edges of the histograms in Fig.~\ref{xF_distr}
and \ref{y_distr} corresponding to the borders of phase space in terms of 
variable $x_F$ and $y$ where the asymmetry reach maximal values.

   We would like cordially to thank M.~Medinnis for his inspiration and 
support of these studies, useful discussions and careful reading of the 
manuscript.  
   We are also grateful to T.~Sjostrand and J.~Carvalho for stimulating 
information and discussions.


\end{document}